# Photo-excitation production of medically interesting isomers using high-intensity γ-ray source

Wan-ting Pan[1], Hao-yang Lan[1], Zhi-guo Ma[1], Zhi-chao Zhu[1] and Wen Luo[1*]

[1]School of Nuclear Science and Technology, University of South China, 421001, Hengyang, P.R. China

[*]Corresponding author: wenluo-ok@163.com

**Abstract:**

Photon-induced nuclear excitation (i.e. photo-excitation) can be used for production of nuclear isomers, which have potential applications in astrophysics, energy storing, and medical diagnosis and treatment. This paper presents a feasibility study on production of four nuclear isomers ($^{99m}$Tc, $^{103m}$Rh and $^{113m, 115m}$In) using high-intensity γ-ray source based on laser-electron Compton scattering (LCS), for use in the medical diagnosis and treatment. The decay properties and the medical applications of these nuclear isomers were reviewed. The cross-section curves, simulated yields and activity of product of each photo-excitation process were calculated. The cutoff energy of LCS γ-ray beam is optimized by adjusting the electron energy in order to maximize the yields as well as the activities of photo-excitation products. It is found that the achievable activity of above-mentioned isomers can exceed 10 mCi for 6-hour target irradiation at an intensity of the order of $10^{13}$ γ/s. Such magnitude of activity satisfies the dose requirement of medical diagnosis. Our simulation results suggest the prospect of producing medically interesting isomers with photo-excitation using the state-of-art LCS γ-ray beam facility.

**Keywords:** photo-excitation; nuclear isomer; medical application; activity; laser-electron Compton scattering (LCS)

## 1. Introduction

A radionuclide which decays by emission of γ rays and/or conversion electrons to reach ground state is called nuclear isomer. In most cases, nuclear isomers emit only gamma radiation and no beta radiation, offering the advantage of low radiation dose for patient investigations [1]. Besides offering great potential in the field of astrophysics and energy storing, nuclear isomers are particularly suitable for various applications in nuclear medicine if they can be produced with high (specific) activity [2]. In nuclear medicine, the present mostly used isomer is $^{99m}$Tc ($T_{1/2}$ = 6.01 hr). Its 140.5 keV γ ray is ideal for single photon emission computer tomography (SPECT) imaging. Other isomers, such as isomers $^{113m, 115m}$In, for example, are used in cancer treatments [3]. Since the high multipole order of the depopulating transitions, most of the decays occur via internal conversion, with subsequent cascade emission of Auger electrons that can be used to kill various cancer cells. There are also a few isomers, including $^{34m}$Cl and $^{52m}$Mn [4, 5], which decay by positron emission and electron capture and are suitable for positron emission tomography imaging.

The radionuclide generator scheme plays a major role in production of many of the known isomers including $^{99m}$Tc and $^{81m}$Kr, which have been widely used in nuclear medicine [6]. In such scheme, a metastable daughter (isomer) can be separated from its longer-lived parent by an effective radiochemical process, such as simple ion exchange process. This scheme promotes the direct use of the known isomers with relatively short half-lives in medical applications. Currently thermal neutron-induced fission and charged particle reactions with protons are the main production routes for radionuclide generators, which rely largely on either research reactors or cyclotrons [7, 8]. However, most of the research reactors used for medical isotopes production are more than 50 years old and approaching the end of their useful lives [9]. Moreover, the production of required parent radionuclide is accompanied by yielding considerable quantities of long-lived radioactive waste. Cyclotrons generate a smaller amount of radioactive waste in comparison to research reactors, but the radionuclide production is very limited depending on installed beam energy. Owing to the limited number of production facilities available and the increasing global demand for medical isotopes, the shortage in medical isotopes, particularly the $^{99}$Mo/$^{99m}$Tc generator, has reached crisis level in the past years [9, 10]. With a serious shortage of medical isotopes looming, worldwide researches are exploring ways to make them without nuclear reactors [11, 12]. These initiatives include the investigation of innovative ways of production and the application of new radioisotopes in both imaging and therapy [13,14]. Recently, photo-production of $^{99m}$Tc with the nuclear resonance fluorescence (NRF) has been proposed by recycling hazardous $^{99}$Tc [15].

In this study, we present a feasibility study for producing longer-lived nuclear isomers for medical usage. This can be achieved by bombarding isotopic target with energetic γ-ray beam to induce photo-excitation process, namely (γ, γ') reaction [16], hence producing nuclear isomer of interest. Laser-electron Compton scattering (LCS) approach is employed to produce such γ-ray beams with variable energy. We reviewed briefly the decay properties of selected nuclear isomers, $^{99m}$Tc, $^{103m}$Rh and $^{113m, 115m}$In and their potential applications in nuclear medicine. We then calculated (γ, γ') reaction cross-section curves and simulated the activities and production yields for each photo-excitation process. In these simulations, we optimized the LCS γ-ray beam energy to maximize the isomer production. It is found that the achievable activity of nuclear isomers can exceed 10 mCi when each of isotopic targets is irradiated at an intensity of the order of $10^{13}$ γ/s for 6 hrs. Such magnitude of activity satisfies the dose requirement of medical imaging, indicating the prospect of producing medically interesting isomers with photo-excitation approach.

## 2. Nuclear isomer via the (γ, γ') reaction

Longer-lived nuclear isomers are of interest for various applications in nuclear medicine if they can be produced with high (specific) activity. Table 1 shows a selection of such isomers that can be produced properly through (γ, γ') reactions. Most usual production methods, e.g., via (n, γ) reactions on $A - 1$ target isotopes, result in relatively low activity since the dominant part of the production proceeds directly to the nuclear ground state that has a nuclear spin closer to that of the $A - 1$ target isotopes. Meanwhile, (γ, n) reactions allow producing neutron-deficient isotopes when LCS γ-ray beams have sufficient flux density [2]. We select four nuclear isomers, $^{99m}$Tc, $^{103m}$Rh, and $^{113m, 115m}$In, which is suitable for being produced through (γ, γ') reactions considering the target availability. Particularly, $^{99}$Tc is independently produced in the reactors and constitutes ~0.1% of total spent fuel inventory [15]. However, they are not able to be produced by (γ, n) reactions due to those unstable $A + 1$ target isotopes. In the following the decay properties of these nuclear isomers are briefly reviewed, together with their potentially medical applications.

Table 1. The decay properties of four nuclear isomers selected and the target isotopes required for

(γ, γ') reactions leading to these isomers.

| Isomer | $T_{1/2}$ (hr) | Exc. energy (keV) | Target isotope | Nat. abundance or half-life | Decay model | Medical application |
|---|---|---|---|---|---|---|
| $^{99m}$Tc | 6.01 | 142.7 | $^{99}$Tc | $2.1 \times 10^5$ a | IT (99.9%) | SPECT |
| $^{103m}$Rh | 0.94 | 39.8 | $^{103}$Rh | 100% | IT (100%) | Radiotherapy |
| $^{113m}$In | 1.66 | 391.7 | $^{113}$In | 4.3% | IT (100%) | SPECT; Radiotherapy |
| $^{115m}$In | 4.49 | 336.2 | $^{115}$In | 95.7% | IT (95%) | SPECT; Radiotherapy |

## 2.1 $^{99m}$Tc

$^{99m}$Tc ($T_{1/2}$ = 6.01 hr) has been widely used for SPECT imaging. With a relatively short half-life and the quasi-absence of beta particles, the radiation dose to the patient is low. A patient can be injected with small amount of $^{99m}$Tc and within 24 hours almost 94% of the injected radionuclide would have decayed and left the body. $^{99m}$Tc can be chemically bound to small molecule ligands and proteins, which are concentrated in specific organ tissues after injection into the body, such as $^{99m}$Tc-HMPAO, $^{99m}$Tc-ECD and $^{99m}$Tc-DTPA [17].

## 2.2 $^{103m}$Rh

$^{103m}$Rh ($T_{1/2}$ = 56.1 min) is a key example of an Auger-emitting candidate. It decays via isomeric transition to stable $^{103}$Rh, and gives rise to emission of low-energy Auger electrons [18]. The 40 keV isometric decay energy is totally converted in the electronic shells for the stable $^{103}$Rh with no measurable γ rays and results in a "shower" of low-energy electrons and X rays [19]. Due to its attractive decay and transition properties, $^{103m}$Rh is expected to exhibit high cytotoxicity and is an isotope of interest for radioimmunotherapy and targeted therapy.

## 2.3 $^{113m}$In

$^{113m}$In ($T_{1/2}$ = 99.5 min) emits a monochromatic photon spectrum with an energy of 391.7 keV. The γ-emitting isomer $^{113m}$In can be incorporated as the radioactive label for a series of compounds with which the brain, kidney, lungs, liver-spleen, and blood pools can be imaged [20]. $^{113m}$In is also an Auger-electron emitter used for targeted therapy [21]. The advantage of using such isomer lies in the increased patient doses which are permitted within the framework of radiation safety.

## 2.4 $^{115m}$In

$^{115m}$In ($T_{1/2}$ = 4.49 hr) is considered to be very potential for molecule labeling purposes in order to obtain suitable preparation for nuclear biology and medical

applications. It decays by isomeric transition emitting γ-ray of 336 keV (~95%) and by β⁻ decay emitting electrons of 0.86 MeV (~5%). Although the electron emission has insignificant contribution to radiation dose, it is strong enough to kill cancer cells so that it can be used for cancer therapy [22]. Radioisotope $^{115m}$In is comparable to the most popular medical diagnostic radioisotope $^{99m}$Tc but with a simpler chemical property as it has only one prominent oxidation state in aqueous solution [23].

## 3. Simulation results

In this section, we first present the simulation of LCS γ-ray beam generation and the modelling of photo-excitation cross sections for four isomers shown in Table 1. We then discuss the isomer production as a function of irradiation time, γ-ray beam energy and target geometry. In the following simulations, the LCS γ-ray beam flux is fixed to $10^{13}$ γ/s, which is obtained at the interaction point (IP) occurring the LCS process; the target used for irradiation has a radius of 2 mm and a thickness of 5 cm by default.

### 3.1 LCS γ-ray beam generation

To induce efficiently the photo-excitation process, a high-brightness γ-ray beam is required, such as Bremsstrahlung beam [24] and LCS γ-ray beam [25]. The LCS γ-ray beam is produced by incoherent Compton back-scattering of laser light from brilliant high-energy electron bunches, in which the laser photons take substantial energy from electrons and then boost successfully into the higher X or γ-ray beam energy region. The energy $E_\gamma$ of the LCS γ-ray beam can be derived from the conservation of energy and momentum and can be expressed as:

$$E_\gamma = \frac{E_L(1-\beta\cos\theta_1)}{1-\beta\cos\theta+E_L/E_e[1-\cos(\theta_1-\theta)]}, \quad (1)$$

where $E_e$ and $E_L$ are the energies of incident electron and laser photon, respectively, $\beta = \sqrt{1-\gamma^{-2}}$ is the electron velocity in terms of speed of light, $\gamma$ is the Lorentz factor of the electron beam (*e*-beam), and $\theta_1$ and $\theta$ are the laser incident angle and the scattering angle with respect to the direction of the *e*-beam, respectively. Considering the head-on collision geometry, the energy $E_\gamma$ can be simplified by [26]

$$E_\gamma = \frac{4\gamma^2 E_L}{1+\gamma^2\theta^2+4\gamma^2 E_L/E_e}. \quad (2)$$

It is shown that the cutoff energy of the scattered photons is proportional to the square of the energy $E_e$ in the linear LCS regime. In order to produce a high-energy LCS γ-ray beam, an energetic *e*-beam should be employed and the scattering angle $\theta \sim 1/\gamma$ should be very small. There are a few of LCS facilities with which very brilliant, intense γ-ray beam can be produced. The Extreme Light Infrastructure - Nuclear Physics (ELI-NP) will construct a γ-ray beam system called variable-energy gamma ray system that will be able to provide γ-ray beam with energies up to 19.5 MeV and spectral photon density of around $10^4$ s$^{-1}$ eV$^{-1}$, which is very suitable for exploring different fields in photonuclear physics, such as NRF, photonuclear reactions, and studies of photo-excitations [27].

Combining with the Geant4 toolkit, a 4D (three-dimensional time and frequency domain) Monte Carlo laser-Compton scattering simulation (MCLCSS) code [28] is used to model the MeV-Class γ ray production and the following transportation. In the simulations we used a circularly polarized laser with a wavelength of 515 nm and wavelength bandwidth of 0.05%, scattering on a relativistic *e*-beam with an energy spread of 0.04% at a laser incident angle of $\theta_1 = 172.5°$. We then diagnosed the γ beam characteristics when the LCS γ-ray beam arrive at the surface of the irradiation target. Figure 1(a) shows the simulated spectral and spatial distributions of LCS γ-ray beam, which are required for the following calculation of the activity of isomers. The spectrum has a sharp edge on high-energy cutoff which is determined by the bandwidths of incident electron and laser photon beams according to Eq. (2). We see that the scattered photons with high energies are concentrated around the center ($\theta = 0$), while the low-energy photons are distributed away from the center. Such a relation, in principle, allows one to obtain readily a quasi-monoenergetic LCS γ-ray beam by a simple geometrical collimation technique.

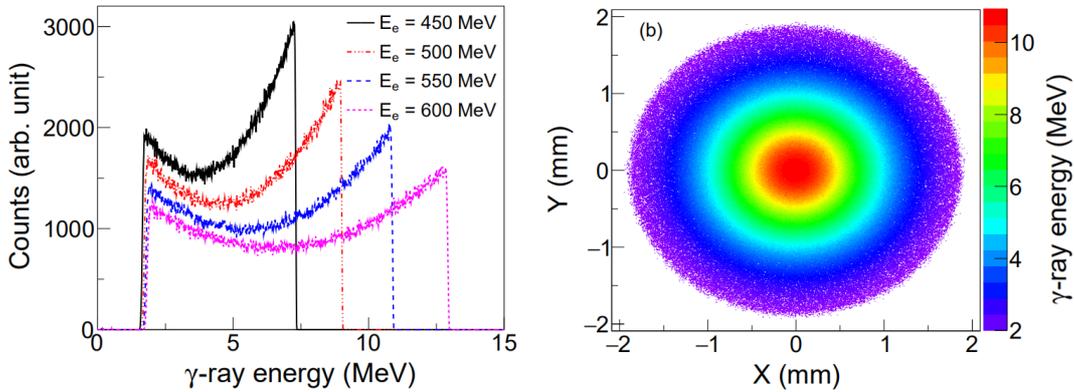

Fig. 1 (a) The simulated γ-ray beam spectral distributions for different $E_e$, and (b) spatial pattern for fixed $E_e$ = 550 MeV. The diagnosed plane is the surface of irradiation target with a radius of 2 mm, which is located 1.5 m downstream of the IP.

## 3.2 Photo-excitation cross section

Since scarce data is currently available for reactions induced on excited states, cross sections for excitation of nuclear isomers in (γ, γ') reactions, $\sigma(E_\gamma)$, have been calculated with TALYS. TALYS is a software package for the simulation of nuclear reactions, which provides a complete description of all reaction channels and observables, and many state-of-the-art nuclear models covering all the main reaction mechanisms encountered in light particles induced nuclear reactions are included [29]. TALYS is designed to calculate the total and partial cross sections, the residual and isomer production cross section, the discrete and continuum γ-ray production cross sections, etc. For the (γ, γ') reactions on four target nuclei, the calculations of the cross sections are performed with TALYS 1.6, which generate the results illustrated in Fig. 2. The nuclear structure ingredients used for the TALYS computations are explicitly presented in [30]. It can be inferred from Fig. 2 that the peak values of $\sigma(E_\gamma)$ have magnitude of the order of mb. As the γ-ray energy increases, the production cross sections decrease rapidly. Specifically, the $^{103m}$Rh has the highest production cross section, followed by the $^{113m,115m}$In isomers. Similar to the beam parameters, the curves of $\sigma(E_\gamma)$ are also necessary for the prediction of the activity of isomers, as discussed later.

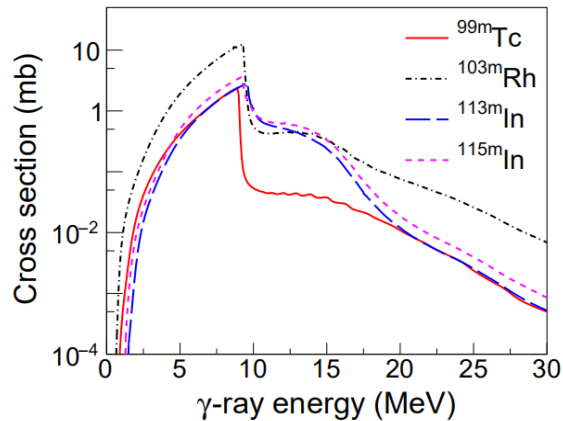

Fig. 2 Isomer production cross-section via (γ, γ') reaction on the target isotopes listed in Table 1.

## 3.3 Production of medically interesting isomers

Considering an isotopic target cylindrically irradiated by the LCS γ-ray beam, the production rate $P$ of isomer of interest induced by photo-excitation process can be expressed as [31]:

$$P = \int_{E_{th}}^{E_{max}} \int_0^R \frac{\sigma(E_\gamma)}{\mu(E_\gamma)} \rho I(E,r)[1 - \exp(-\mu(E_\gamma) \cdot L)] dEdr. \qquad (3)$$

Here, $E_{th}$ is the threshold energy of photo-excitation reaction, $E_{max}$ is the cutoff energy of incident photons, $R$ and $L$ are the radius and length of the isotopic target $I(E_\gamma, r)$ is the flux density of γ-ray beam irradiating the front surface of the target, and $\rho$ is number density of the target nuclei, and $\mu(E_\gamma)$ is the linear attenuation coefficient. Although the isomer is formed at the production rate $P$, itself still decays at the rate $\lambda N_i(t)$, and hence the net production rate is given by:

$$\frac{dN_i(t)}{dt} = P - \lambda N_i(t), \qquad (4)$$

where $\lambda = \ln 2/\tau$ is the decay constant of the isomer of interest with a half-life of $\tau$. By solving Eq. (2) and substituting the initial conditions, one could obtain the isomer activity as a function of irradiation time $t_{irr}$:

$$A_0 = P(1 - e^{-\lambda t_{irr}}). \qquad (5)$$

One can see that the isomer activity increases with irradiation time. After a few of half-life irradiation interval, the isomer activity reaches saturation value. When the irradiation stops, it then decays exponentially following $A_i(t) = A_0 e^{-\lambda t}$. In order to evaluate the production of isomer of interest, the amount of isomer decayed within a certain time interval can be obtained with:

$$S = \int_0^t A_i(t) dt = \frac{P}{\lambda}(1 - e^{-\lambda t_{irr}})(1 - e^{-\lambda t}) \qquad (6)$$

In this study, we used a data-based Monto Carlo simulation program [31] to calculate the production rate $P$ and then to obtain the activity of medically interesting isomers. The spectral and spatial patterns of the γ-ray beam and the photo-excitation cross sections are chosen as input data, which also includes the attenuation coefficient of γ-photons inside target [32]. Figure 3 shows the activity of selected four isomers as a function of $t_{irr}$. The isomer activity increases notably with irradiation time and then reaches a saturation value after more than 5 times the half-life irradiation interval. Furthermore, the isomer activity is closed related to the convolution between the γ-ray beam spectrum (see Fig. 1) and the production cross-section (see Fig. 2). It is seen that the achievable activity can exceed 10 mCi when the target is irradiated for 6 hours.

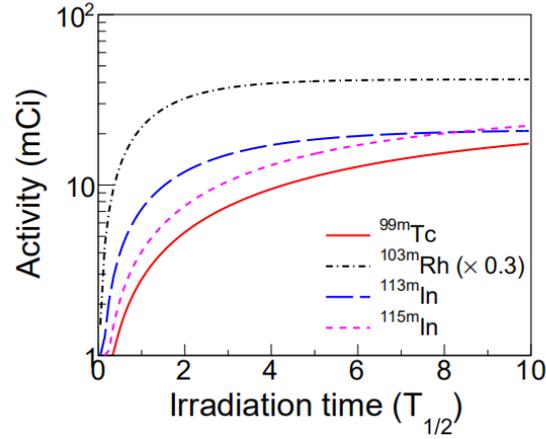

Fig. 3 Isomer activity as a function of irradiation time. The LCS γ-ray beam energy used for calculation is 9.5 MeV.

Considering that the photo-excitation cross section is fixed for production of specific isomer, one may optimize the isomer activity by changing the energy spectrum of incident γ-ray beam. Figure 4 shows the isomer activity as a function of the cutoff energy of LCS γ-ray beam. It is found that for the given target dimension, the LCS γ-ray beam with a cutoff energy of 9.1 MeV could result in peak activity for $^{99m}$Tc. The isomers $^{103m}$Rh and $^{113m, 115m}$In have different situations. Their peak activities occur at the cutoff energy of ~9.5 MeV. These results show that the matching the LCS γ-ray beam spectrum with a peaked $\sigma(E_\gamma)$ by using an appropriate γ-ray beam may provide an effective approach to producing a maximum activity.

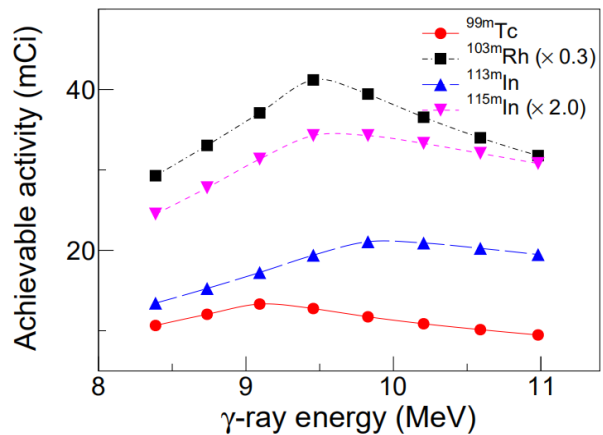

Fig. 4 Achievable activity as a function of LCS γ-ray beam energy with 6-hour irradiation

In order to realize a better understanding on the photo-excitation process induced by high-intensity LCS γ-ray beam, we performed Geant4 simulations [28] and diagnosed the spatial pattern of the product isomer, such as $^{99m}$Tc, under the condition of optimized beam energy. It is seen in Fig. 5 that the intensity of $^{99m}$Tc drops along both radial and longitudinal directions and its peak intensity is found in the center of the target isotope. This effect is correlated well with the spatial pattern of the incident γ-ray beam, and is independent on the shape of the $\sigma(E_\gamma)$. Since the low-energy photons are distributed away from the center of the target isotope and have relatively weak penetrability, the intensity of $^{99m}$Tc shows a cone-shaped distribution in the X-Z plane. One can see that spatial pattern could provide the knowledge to choose approximate target dimension. In this case, a $^{99}$Tc target with less than 1.0 mm radius may match the irradiation requirement.

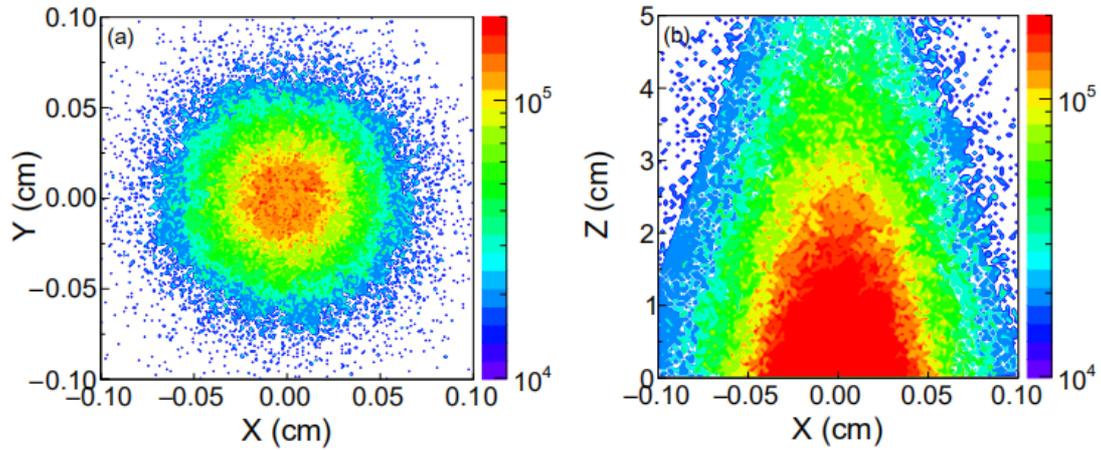

Fig. 5 Projective distributions of production position of $^{99m}$Tc in the X-Y plane and X-Z plane. The propagation of the incident γ-ray beam is along the Z-axis direction.

Figure 6 shows the dependence of isomer activity on target thickness. For $^{99m}$Tc, $^{103m}$Rh and $^{113m,115m}$In isomers, the achievable activity increases with the increasing target thickness. However, when the target is thicker than 5 cm, the increase trend becomes slower due to the significant attenuation of the incident γ-ray beam; particularly, $^{99m}$Tc and $^{103m}$Rh isomers have a saturation activity. It has been shown that the specific activity decreases with the target thickness increased [13]. As a result, one needs to compromise with the target thickness in order to produce isomers of interest with sufficiently high activity as well as specific activity.

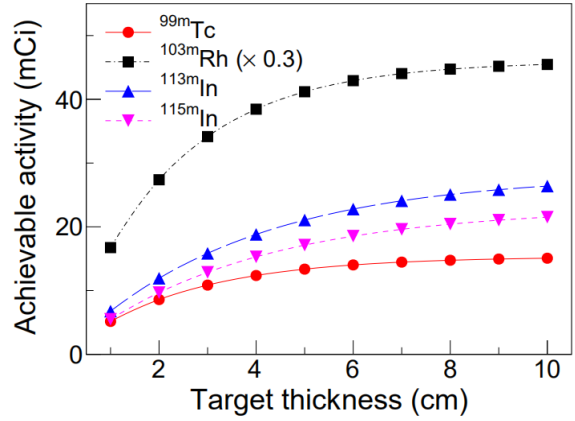

Fig. 6 Achievable activity of medically interesting isomers as a function of the target thickness. The calculations were performed with optimized γ-ray beam energy. The isotopic target has a radius of 2 mm and the irradiation time is 6 hours.

## 4 Discussion

In order to summarize our results, we present in Table 2 the optimized beam energies, $E_{\gamma,opt}$, with respect to the energy $E_{\gamma,peak}$, leading to the peak value of the (γ, γ') cross section $\sigma_{peak}$. It suggests that the value of $E_{\gamma,opt}$ should be slightly higher than the $E_{\gamma,peak}$. Similar phenomenon has also been observed in medical isotope production via (γ, n) reactions [13]. Table 2 also shows the estimates of the achievable activities for four isomers selected. It is found that their achievable activities exceed 10 mCi when each of isotopic targets is irradiated at an intensity of the order of $10^{13}$ γ/s for 6 hrs. Generally, mCi levels of radioactivity are adequate for SPECT imaging, whereas several hundreds of mCi of activity are often required for therapeutic applications [20]. As a result, the proposed approach, together with an intense LCS γ-ray beam, provides unprecedented opportunities for production of sufficient amounts of the isomers for medical imaging.

Table 2 A brief summary of medically interesting isomers that can be produced with intense LCS γ-ray beam. To calculate the isomer activity, each of isotopic targets is irradiated at an intensity of the order of $10^{13}$ γ/s for 6 hrs.

| Isomer | $\sigma_{peak}$ (mb) | $E_{\gamma,peak}$ (MeV) | $E_{\gamma,opt}$ (MeV) | $\mu(E_{\gamma,opt})$ (cm$^{-1}$) | $A$ (mCi) | Amount within 6-hr (Ci) |
|---|---|---|---|---|---|---|
| $^{99m}$Tc | 2.4 | 8.9 | 9.1 | 0.43 | 13.36 | 208.27 |
| $^{103m}$Rh | 12.5 | 9.3 | 9.5 | 0.43 | 137.29 | 662.23 |
| $^{113m}$In | 2.9 | 9.6 | 9.8 | 0.26 | 21.07 | 166.82 |
| $^{115m}$In | 3.6 | 9.2 | 9.5 | 0.25 | 17.13 | 241.26 |

## 5 Conclusion

The $^{99m}$Tc, $^{103m}$Rh, and $^{113m, 115m}$In isomers have suitable decay properties and hence can be used in various medical applications. We have shown that the production, through photo-excitation process, of these four isomers is possible for optimized γ-ray beam energy and target geometry. The cross-section curves, simulated yields and activities of production of each photo-excitation process were calculated. Optimal cutoff energy of LCS γ-ray beam are found to produce four medically interesting isomers with maximal activity. When isotopic target experience 6-hour irradiation at an intensity of the order of $10^{13}$ γ/s, the $^{99m}$Tc, and $^{113m, 115m}$In isomers can be produced with activity of 10 mCi, and the $^{103m}$Rh isomer with activity of 100 mCi. We conclude that the proposed photo-excitation approach, combined with the state-of-art LCS γ-ray beam facility, is suitable for production of $^{99m}$Tc, $^{103m}$Rh, and $^{113m, 115m}$In in sufficient quantities for medicine usage.

## Acknowledgment

This work is supported by the National Natural Science Foundation of China (Grant No. 11675075), and Natural Science Foundation of Hunan Provence, China (Grant No. 2018JJ2315). W.L appreciates the support from the Youth Talent Project of Hunan Province, China (Grant No. 2018RS3096).